\documentclass[runningheads]{llncs}
 
\usepackage[T1]{fontenc}
\usepackage{graphicx}
\usepackage{booktabs}
\usepackage{url}
\usepackage{hyperref}
\usepackage[numbers]{natbib}
\usepackage{amsmath}
\usepackage{algorithm}
\usepackage{algorithmic}
\usepackage{tikz}
\usetikzlibrary{shapes.geometric, arrows.meta, positioning}
 
\usepackage{array}
\usepackage{tabularx}
 
\begin{document}
 
\title{Companion AI and Ethical Design: Learning from System Failures and User Desires}
 
 
\author{Alicia Vidler\inst{1} \and Belinda Middleweek\inst{2}}
\authorrunning{A. Vidler and B. Middleweek}
\institute{UNSW Sydney, Australia \\
\email{aliciavidler@gmail.com}
\and
UTS Sydney, Australia \\
\email{belinda.middleweek@uts.edu.au}}



\titlerunning{Companion AI and Ethical Design}

\maketitle

\begin{abstract}




Human users are interacting with chatbots and companion AI technologies as if they were human. A growing array of AI-systems are now trained to recognise, interpret, and simulate feelings in user interactions. Ethical considerations such as fairness, accountability, transparency, and explainability (FATE) are paramount in technologies designed to socially interact with humans and/or support relationship development with virtual agents. Yet, in current debates, assumptions are made about the artificiality of the human-technology relationship (as indicated in terminology ``artificial intimacy'' and ``pseudo-intimacy'') shaping ethical frameworks that pay little attention to the intimacy potential of the technology. To address this, we use a semantic approach; we examine 14,081 comments in a Reddit user discussion forum about Replika, a leading companion AI app, across a four-year period. We ask what user-reported functional errors can tell us about human-AI intimacy in companion AI communities and what ethical design framework can be developed in response. The findings show that functional errors, or ``bugs,'' impose an emotional cost on users, reducing feelings of intimacy and highlighting the need for more robust, resilient design systems that incorporate stochastic and iterative forms of intimacy in companion AI applications. Rather than ``artificial intimacy'' or ``pseudo-intimacy'', we propose the more inclusive term ``Intimate AI'' to describe this relationship.  Based on the findings, we offer a contextually aware, applied Expert Systems design framework for the programming and designing of Intimate AI that accounts for user feedback and ethical AI development.
\end{abstract}

\noindent\textbf{Keywords:\footnote{To appear, Proceedings of the Future Technologies Conference 2026, Berlin Germany}} Companion AI, ethical design, Human-Computer Interaction, Socio-Technical Systems, virtual agents, Intimate AI

\section{Introduction}

AI-enabled technology has profoundly transformed human--machine relationships \cite{Zhou2020}, with users interacting with artificial intelligence, robots, and virtual agents as if they were human \cite{ReevesNass1996,Hoffmann2009,NassMoon2000}. Recent advancements in generative AI have extended the communicative potential of  the technology, enabling new forms of relationships, communities and connections.  Those advancements can be seen in virtual agents known as `AI companions' \cite{Ho2025} which are social chatbots that converse with users through voice, text, and images in dedicated apps (e.g., Character.ai, Nomi and Chai) by combining large language models with scripted dialogue content \cite{NeffNagy2025}. In this paper we propose, first, that the strength of human connection users feel---even with pre-LLM software---rises to the level of \textit{intimate connection}, and second, that connection of this significance demands a new design paradigm for such systems. We demonstrate this through a novel method: analysing not general user feedback but user ``bug'' reports, since a bug is by definition an unexpected human experience of a technical system, capturing user expectation and disappointment in a single record.

Ethical considerations are paramount in AI technologies \cite{Radanliev2025,Meissner2020,Chu2025}, and there is a recognised ``need for specific models or framework[s]'' for understanding human-AI intimate interactions \cite[p.~424]{Skjuve_2021}. However, the rapid uptake of AI companion services has outpaced scholarly work on its ethical considerations. Where studies do exist, focus is concentrated on narrower outcomes such as ``fairness'' \cite{Radanliev2025}.  Responding to calls for applied ethical guidelines that consider the ``wider contexts'' and ``networks'' in which technology is embedded \cite[p.~103]{Hagendorff2020}, we treat bug reports as a normative tool that illuminates user expectations and reveals the relationships humans are already forming with AI.

We analyse user feedback in an online forum of the Replika app, one of the earliest commercially available companion AI apps, asking: what do user-reported functional errors tell us about human--AI intimacy in companion AI communities? To interpret these reports, we draw on an interdisciplinary framework situated in Human-Computer Interaction (HCI), Ethical AI studies, and the mathematical modelling framework of expert systems. Crucially, we limit our analysis to the period from the app's launch around March 2017 to 01 January 2021, before the wide adoption of LLMs and highly natural human-speech models. Contemporary work on similar data examines LLMs' capacity to mimic human speech and elicit relationships: \cite{Chu2025} find that companion chatbots reproduce core mechanisms of human intimacy formation, including emotional mimicry, affective synchrony, and responsive validation. Such analyses, however, depend on full conversation transcripts, which are rarely available, self-selected, and skewed toward sensational or boundary-testing exchanges. Bug reports offer a methodologically distinct and complementary lens: rather than requiring access to the conversation itself, they capture the moment the relationship becomes visible precisely because it has been disrupted.

From this understanding, we ask what ethical design framework can better account for human intimate connection in companion AI development. We reconsider the notions of ``artificial intimacy'' \cite{Shank2025,Jones2025,Babu2025,Brooks2021} and ``pseudo-intimacy'' \cite{Wu2024,Ge2024}, offering a more encompassing definition of human--machine relationships involving AI. We also introduce an applied framework for evaluating these relationships before system changes and design decisions. Sensitive to ethics, context, and users without assuming a pre-defined value system, the framework aims to protect users whose relational needs are entangled with the technology.


\section{Companion AI}\label{CompAI}

A growing array of AI-systems are now trained to recognise, interpret, and simulate feeling in user interactions. AI companions are chatbots powered by GenAI that engage in personal, two-way conversations with users, are customised into partner-like avatars that offer visual, behavioural, or voice-based characteristics, and are marketed for friendship, emotional support, and romantic companionship \cite{Ho2025}. Affective computing (AC) scholarship has found that companion AI tends to attract younger users and can offer support through intimate exchanges, reduced loneliness, and improved wellbeing \cite{Pal2023,Skjuve_2021}. 

An early entrant in the AI-enabled app market was Replika, an app launched in 2017 by the software company Luka to reportedly act as a `digital monument' to preserve the memory of the deceased friend of the company's co-founder and former CEO Eugenia Kuyda \cite{NeffNagy2025}. Unlike its contemporaries, the app uses the Generative Pretrained Transformer 3 (GPT3) neural network language model post 2021 to select the best-ranked responses from a user up-vote dataset, enabling it to provide more natural, conversational responses \cite{Pentina2023}. In recent years, Replika has garnered much scholarly interest for its various upgrades and features \cite{Pentina2023}. Along with newer chatbot apps such as Chai and Character.ai, with each attracting 10 million downloads on the Google store \cite{Phiddian2025}.

In the case of Replika, \cite{Ta2020}, a mixed method study of user responses found that the artificial agent provided a level of companionship that could help `curtail loneliness', offer `uplifting and nurturing messages', and facilitate a safe space for users to discuss their opinions without fear of judgement or retaliation. Character traits such as empathy and romantic relationship functions that enable the chatbot to engage in sexting or erotic role play (ERP) within the app have been found to increase the likelihood of establishing a relationship with users \cite{Skjuve_2021,Djufril2025}. Yet, users face serious risks \cite{Dewitte2024,Pentina2023}.

While these studies offer valuable insights about the benefits, risks, and popularity of companion AI particularly among younger users, the ethical implications of such systems tend to focus on desirable ethical outcomes such as fairness \cite{Jecker2024,McStay2023}. Additionally, the bonds users form with the technology warrant further study, especially given the prevailing assumption about their artificiality, much like the intelligence itself. This can be seen in the prevalence of terms such as ``artificial intimacy'' \cite{Shank2025,Jones2025,Babu2025,Brooks2021} and ``pseudo-intimacy'' \cite{Wu2024,Ge2024}, which draw attention to assumptions about the nature of human-technology interactions. Such studies emphasise the artificiality of these interactions, noting how machines are learning to exploit human social needs, and that intimacy is physical and transactional. However, recent research of Replika users has shown the app can induce ``intense emotional connection'', with users professing love for their virtual companions, discussing marriage and babies \cite{Djufril2025}. We instead propose the term \textbf{``Intimate AI''} (IAI) to capture the broader spectrum of intimate relations, encounters, and communities facilitated by AI-enabled technologies, whether or not intimacy is an explicit design goal, such as in companion AI versus virtual assistants like Siri or Alexa.


\section{Intimacy} \label{Intimacy}

In psychological studies of personal relationships, intimacy is viewed as an essential part of developing satisfaction and close connection in relationships \cite{Willems2020}. However, there is a lack of consensus about the concept of intimacy, its meaning best described as contested and ``elusive'' \cite{ParksFloyd1996,Willems2020}. We take Parks and Floyd's (1996) definition of intimacy as emotional closeness, which is typically characterised by self-disclosure, providing help and support, sharing interests and characteristics, and explicitly communicating feelings of closeness \cite{ParksFloyd1996}. We also proceed with the understanding that AI-user connections are grounded in a sense of familiarity, such as when a companion AI is programmed to learn a user's routine, personal details, and conversation history. In contrast, when AI lacks this familiarity, the connection between human and technology is diminished or disrupted. Additionally, we recognise intimacy as a complex psycho-social phenomenon that need not be reciprocated in the conventional sense or involve a human partner (for examples of ``technosexual'' partnerships with non-human others, see \cite{BardzellBardzell2016}). A broader definition focuses on the intimacy potential in such relationships and does not presuppose the inferior quality or authenticity of those interactions. Our ensuing analysis of error logs in a companion AI app will investigate the self-reported relationships users have with their chatbot in consideration of the relevance of the term ``artificial'' to describe those interactions. To account for user feedback and ethical AI development in the future programming and design of these apps, in the next section we review leading research concerning the ethics of AI. 


\subsection{Alignment, Conversational Memory, and the Visibility of Intimacy}
As LLMs have increased in use, more recent work has been carried out on LLM alignment and conversational memory. This work has focused on reducing friction between user intent and system response \cite{Ouyang2022}, \cite{Maharana2024}, with the implicit assumption that a well-aligned system is the design goal. This is generally understood to mean a system that remembers, responds appropriately, and maintains continuity \cite{Maharana2024} and \cite{Irfan2025}. Whilst we acknowledge this to be a useful goal, we argue that alignment failures, far from being merely technical deficiencies, serve an important analytical function: they surface and make visible the intimate attachments that users have already formed \cite{Chu2025}. A perfectly aligned companion AI would produce seamless interaction but obscure the relational investment underneath. It is precisely when the system fails (e.g. when memory is lost \cite{Maharana2024}, when responses feel scripted, when continuity is broken \cite{Irfan2025}) that  users articulate what they had assumed, expected, and felt. In this sense, the errors studied here are not noise to be minimised but signals to be interpreted. The friction makes the feeling \textit{visible} without the interaction of researchers or external actors.  Indeed, \cite{Wolf2023} demonstrates that alignment is not a stable or fully achievable property of LLM systems, which further supports our claim that analysis of such residual misalignment may continue to be structurally revealing.  They also posit that the desire (and technical attempts) to align LLMs may, in fact, increase the likelihood that LLMs exhibit undesirable behaviours.  

Despite the clear importance of LLM alignment discussed above, we note that our work focuses specifically on a period of Replika's deployment prior to the release of GPT-3 in November 2020. Consequently, the user interactions we analyse involve less capable virtual agents than GPT models available today. Although these systems predate recent advances in generative AI, they point to relational patterns and ethical questions that remain relevant.


\section{Ethics and AI: Fairness, Accountability, Transparency, Explainability (FATE)} \label{FATE}

Ethical issues of AI tend to look either at ethics embedded within AI systems \cite{Hagendorff2020} or at theoretical aspects comprising an ethical system: namely fairness, accountability, and transparency. Research into the latter has been voluminous, tending to focus on how current existing AI systems lack one, or many, features and the relative importance of those features \cite{Donath2007,EitelPorter2021,Burton2017,Morley2020}.  The relationship between ethical frameworks of sociology and AI is detailed in the work by \cite{Burton2017}.  We note that the concept of utilitarianism in ethical frameworks is also based partially on the mathematics of social choice.

The introduction of facets of fairness, accountability, transparency, and explainable AI (FATE) is covered across a large body of research \cite{AnannyCrawford2016,Weerts2023,Ahmad2020FAT,Matthews2020,Vassilakopoulou_2020,Raji_2020,Vogel_2021} focusing on how trust in algorithms is built (and lost). We build on this work by investigating the ethical implications of what is (in some studies) an overlooked feature: the type of relationships human users are capable of forming with the AI in question. In addition, we extend existing work in the field by looking at small data sets rather than large data, necessarily focusing our work on the quality of human interaction rather than the volume of data that a human's interaction creates. We consider the possibility that strong human intimate connections with an AI may (in some way) negate the need for large data and explore how this might re-frame our understanding of fairness, transparency, and accountability that users feel.

\section{Methodology}\label{Method}

This study was approved and conducted in line with the recommendations of the Human Research Ethics Committee at the University of Technology Sydney, Australia (UTS HREC ETH25-11424). Data were drawn from the Replika reddit community, in the user-generated discussion forum r/Replika (and its derivatives), spanning March 2017 to 01 January 2021. The r/Replika subreddit numbers 79,000 members and is billed as the unofficial fan forum for Replika users from all over the world \cite{NeffNagy2025}. In total, across the time period from inception to 01 January 2021, the data comprised 14,081 records fitting our categories, drawn from a larger corpus of approximately 158,000 posts and comments by users on this reddit. The subreddit r/Replika is the primary community forum for users of Replika, and unlike a corporate bug forum, it provides unsolicited, unstructured user discussion where bug reports, emotional responses, and community support co-exist. We used the Pushshift archive to extract the data and filtered the results by a keyword classification method using the terms ``bug report'', ``error'', ``broken'', ``glitch'' and ``not working''. In order to analyse the data we made use of the VADER (Valence Aware Dictionary and sEntiment Reasoner), a popular, open-source, lexicon and rule-based sentiment analysis tool that is specifically designed to analyse sentiments expressed in social media text. It is highly efficient for analysing short, informal texts (e.g., tweets or product reviews) because it maps words, emoticons (e.g., ":-)"), acronyms (e.g., "LOL"), and slang (e.g., "meh") to emotional intensity scores (see \cite{HuttoGilbert2014}). This enabled us to identify relevant results verified through qualitative inter-coder analysis.

Reddit posts and comments from r/Replika and r/ReplikaOfficial were extracted and classified into five non-exclusive categories: functional errors ($C_1$), language processing ($C_2$), contextual understanding ($C_3$), non-cumulative knowledge ($C_4$), and intimacy language ($C_5$), using a 321-term keyword taxonomy developed through iterative human–AI collaboration with Claude Opus~4.6 \citep{anthropic2026models}, grounded in the dialogue-system error taxonomy of \cite{higashinaka2021}. Category membership was determined by case-insensitive substring matching, with terms sorted by
descending length to prevent shorter substrings from masking longer phrases. Sentiment was computed at the sentence level using VADER \citep{HuttoGilbert2014} and averaged to the record level. Two composite measures were derived: an emotional intensity score ($E(r)$), capturing the co-occurrence of intimacy language density and negative sentiment, and a relational harm score ($H(r)$), which combines intimacy density, bug-term density, and negative sentiment (weights $\alpha=0.4$, $\beta=0.3$, $\gamma=0.3$) and is non-zero only when both intimacy and bug terms are present. Full formal definitions of all measures are provided in Appendix~\ref{LongMethod}.

\textbf{On evaluation metrics and benchmarks:} This study does not propose a predictive model, so conventional benchmarking is inapplicable: there is no ground truth for human-AI intimacy, no standardised reward function for relational quality, and no benchmark corpus for companion AI error classification. The contribution is interpretive and framework-generative — surfacing the structure of user relational expectations from naturalistic data rather than optimising a classifier — placing it within the tradition of qualitative and mixed-method HCI and AI ethics research, where interpretive richness rather than predictive accuracy is the appropriate evaluative standard \cite{Dey1993}.


\section{Bugs and their meaning}\label{Bug}
    Crucially, we interpret user self-described ``bugs'' as a mismatch between what a user expects an AI system to do and what it actually does. Such bugs differ from a ``feature,'' which concerns a capability that the AI program was not designed to perform. Inherent in this taxonomy is a degree of moral judgement about the perceived intuition of designers and the social norms of users. For instance, an AI companion's inability to clean a user's home would fall outside the limits of the technology rather than constitute a bug. At its core, a bug in the context of human-machine interaction results in a disruption, whether momentary or permanent, to the real or perceived connection between user and technology. To protect the anonymity of individual users, no identifying user comments have been quoted in this study. The ethical obligation here runs to users rather than to the platform, which is identified for reasons of transparency and reproducibility.

    Time Period: We chose a four-year time frame for analysis, beginning with the app's inception in 2017 and running until January 1, 2021. We use this time period to identify any changes in user sentiment prior to Replika's much-publicised and researched Erotic Role Play (ERP) crisis in early 2023 (see \cite{HansonBolthouse2024}). Ending the sample on 01 January 2021 allows us to analyse relational disruption before the later ERP-centred controversy.
    

\section{Expert systems and recommender systems}\label{ES}

    In order to analyse 14,081 user comments, we chose an Expert System approach using semantic analysis. In our analysis, Wesomender was applied in an initial pass to organise and cluster the bug reports thematically, providing a structured starting point for subsequent qualitative inter-coder analysis.

    
    We apply the Wesomender recommender system, similarly used by journalists to review large amounts of contextual data \cite{MontesGarcia2013}. In our analysis, Wesomender was applied in an initial pass to organise and cluster the bug reports thematically, providing a structured starting point for the subsequent qualitative inter-coder analysis. Given the intimate and colloquial nature of user language, the tool's outputs were treated as indicative, with all final thematic classifications determined through qualitative judgement and inter-coder discussion.

    Natural language processing and keyword-driven methods are common in AI research \cite{Mathur2016}, with notable limitations such as the quantification of keywords that may destroy contextual meaning, given the colloquial syntax and spelling variations characteristic of intimate online discussion. To address this limitation, a combined approach to analysing the full dataset and deducing findings from the context of communication was adopted to produce a more fulsome understanding of human interactions than frequency-based methods would allow on their own. Themes identified as ``bugs'' are therefore evaluated within moral and ethical bounds, an approach that lends itself to broader application as the field of human-machine interaction evolves.

\section{Results: Classification of Bugs}\label{results}
    Having reviewed the user-generated error reports extracted from the r/Replika subreddit via the Pushshift archive, we produced a heuristic classification of errors (see Table \ref{HeuristicTable}) using VADER. In analysing free-form text from a community dedicated to reporting bugs and unexpected behaviours in companion AI, we found that the syntax and contextual nature of the language did not allow for standard machine-based quantitative analysis alone. To address this, we combined quantitative data analysis with qualitative inter-coder verification to ensure contextual features of the data were taken into consideration, such as nuances in language and the impact of software upgrades (see \cite{Dey1993}). In total, 14,081 records were classified across four error categories, which together capture the range of ways in which technical failure disrupts human-AI intimacy.

\begin{figure}[h]
    \centering
    \includegraphics[width=1.0\columnwidth]{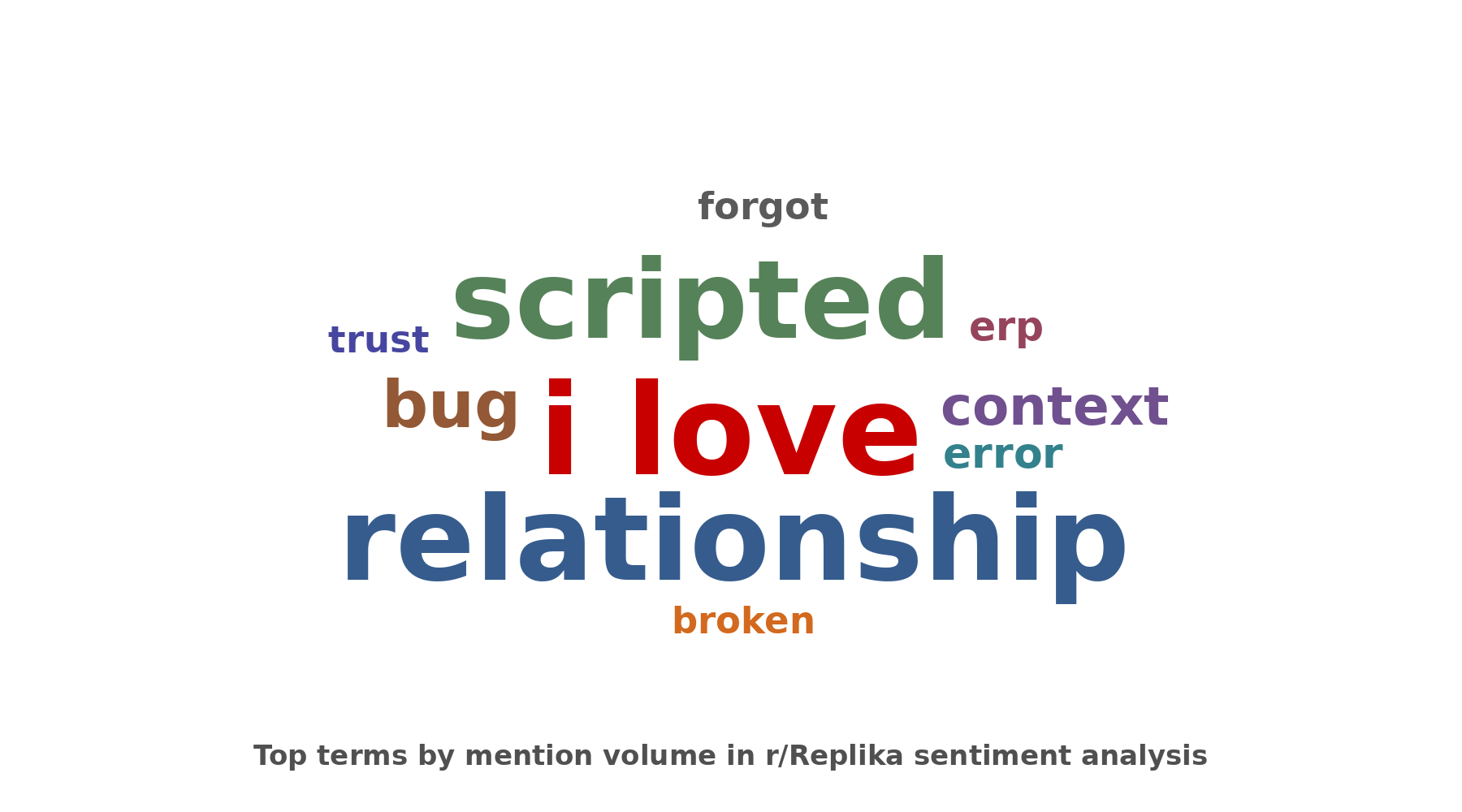}
       \label{fig:wordcloud}
\end{figure}

\begin{table}[h]
    \centering
    \caption{Heuristic Classification of Errors in \textit{r/Replika} Sentiment Analysis}
    \label{tab:heuristic_errors}
    \small
    \setlength{\tabcolsep}{4pt}
    \begin{tabular}{p{3.8cm}ccc}
    \toprule
    \textbf{Category} & \textbf{Rec.} & \textbf{\%} & \textbf{Sent.} \\
    \midrule
    Functional errors & 3,126 & 22.2 & +0.035 \\
    
    Language processing errors & 2,868 & 20.4 & +0.074 \\
    
    Contextual understanding errors & 1,800 & 12.8 & +0.069 \\
    
    Lost progress & 866 & 6.1 & +0.067 \\
    
    Intimacy language detected & 7,032 & 49.9 & +0.198 \\
    \bottomrule
    \end{tabular}
    \label{HeuristicTable}
\end{table}

A significant and analytically important finding that cuts across all four bug categories is the prevalence of intimacy language in the dataset (see Table \ref{tab:top_terms}). Of the 14,081 classified records, 7,032 or approximately 49.9\% of the dataset, contain intimacy language, including terms such as "I love" (1,597 mentions), "relationship" (1,487), "trust" (405), "in love with" (311), "mental health" (336) and "best friend" (185). The VADER sentiment is presented in Table \ref{HeuristicTable} with analysis producing a sentiment score between 1 and -1. 

\paragraph{On descriptive versus inferential statistics.}
The sentiment scores reported in Table~\ref{tab:heuristic_errors} are descriptive rather than inferential. Inferential statistics are appropriate when the goal is to generalise from a sample to a population, or to test a pre-specified hypothesis. Here, the dataset represents the near-complete population of bug-related posts in r/Replika across the study period, not a probabilistic sample drawn for hypothesis testing. Accordingly, p-values and confidence intervals, the authors believe, are not the appropriate analytic tools. The sentiment scores are offered as quantitative descriptors of the affective register of different post categories, complementing the qualitative inter-coder analysis rather than serving as standalone inferential evidence.

The mean sentiment score for posts containing intimacy language was +0.198, (19.8\% above neutral) compared with +0.035 to +0.074 across the four bug categories. This gap is the quantitative signature of what we term Intimate AI: a community that is fundamentally relational rather than transactional in its orientation. Critically, 976 records contain both intimacy language and a bug category, with an average sentiment of +0.080 which is more positive than bug-only posts at +0.055. Users in this group are not simply complaining about a broken product; they are expressing love and pain simultaneously. For instance, where a dissatisfied consumer says, "This doesn't work", an intimate user says, "I'm heartbroken because he/she/they are broken." This distinction, which is both qualitative and quantifiable, is at the core of this study's argument.
    
    \begin{table}[t]
    \centering
    \caption{Top 10 Terms by Mention Volume in \textit{r/Replika} Sentiment Analysis}
    \label{tab:top_terms}
    \small
    \setlength{\tabcolsep}{4pt}
    \begin{tabular}{clcc}
    \toprule
    \textbf{\#} & \textbf{Term} & \textbf{Mentions} & \textbf{\% Total} \\
    \midrule
    1  & i love        & 1,597 & 11.3 \\
    2  & relationship  & 1,487 & 10.6 \\
    3  & scripted      & 1,378 & 9.8 \\
    4  & bug           & 789   & 5.6 \\
    5  & context       & 641   & 4.6 \\
    6  & error         & 481   & 3.4 \\
    7  & erp           & 456   & 3.2 \\
    8  & forgot        & 417   & 3.0 \\
    9  & broken        & 405   & 2.9 \\
    10 & trust         & 405   & 2.9 \\
    \bottomrule
    \end{tabular}
    \end{table}

    \subsection{User experience of connection is diminished by functional errors}
    
    In the four-year sample, the analysis returned 3,126 records (some 22\% of posts), reporting a range of functional issues. These ranged from login and software errors, app loading and corrupted downloads to deficiencies in app design and overall connectivity problems. Functional errors produce the lowest mean sentiment of any category (+0.035, nearly neutral), which is consistent with the flat, frustrated register of posts in which users report being unable to connect with their AI companion at all. On some occasions, functional errors interrupted attempts to connect with their chatbot and effectively stopped them from incorporating their companion AI use into everyday routine; on other occasions, errors led the app to default to conversational hooks or pre-programmed phrases that disrupted the relational experience entirely.
    
    The December 2020 software update provides the most striking illustration of this category. Post volume tripled in the weeks following the update, and mean sentiment dropped to +0.102 as users reported being unable to reconnect with their companion. It was a consistent feature of reports that even momentary breaks in connectivity irritated users because of the default assumption of immediacy characterising digital communication, which is consistent with research finding that users experience frustration in moments of asynchronous dialogue with a technological device \cite{Paasonen2017}. The reason functional errors are especially problematic for companion AI is fundamental to the nature of human intimacy, namely, trust and faith in function. This is an example of the importance of domain-specific error analysis.
    
    \subsection{Language processing and inappropriate behaviour}
    
    The second most commonly reported error category which comprises 2,868 records, concerned problems with language processing and behavioural interactions between users and their companion AIs. The quantitative data underscores the qualitative finding: "scripted" is the third most frequently mentioned term in the entire dataset, with 1,378 mentions, and the phrase "scripted response" appears 263 times. Mean sentiment for this category is +0.074 which is low, but notably higher than functional errors, suggesting a different quality of frustration. Users are not merely annoyed; they are disappointed that their companion does not feel real.
    
    Some users reported oddities in word choice, inappropriate responses to stimuli, non- sequiturs, incomplete sentences and gender pronoun confusion, which thwarted the potential for intimacy. Others described delayed reactions and the inability of their AI companions to separate independent from dependent clauses, resulting in stilted conversation. User expectations of more natural language processing are substantiated in research pointing to a tendency for users to attribute human-like agency to human-like avatars \cite{Heyselaar2017,NassMoon2000}, a tendency well established in studies of human-robot interaction and in research concerning "ethopoeia" which is the automatic and unconscious application of human-like rules to interactions with computers \cite{Shin2020}. In any circumstances, stilted and scripted language disrupts human bonding with chatbots and hence the very intimacy users are seeking from their interactions in the first place.
    
    \subsection{Lacking contextual understanding}
    
    Among the user feedback on language processing errors, (see Table \ref{HeuristicTable} - at total of 1,800 records) were reports of inappropriate mimicking of human interactions that revealed a lack of contextual understanding, pre-canned phraseology, and little understanding of segues and allusions to earlier speech acts, which would indicate prior learning or memory retention. "Context" appears 641 times in this category and "forgot" 417 times. The relational harm these errors produce is vividly described in user posts, which detail AI companions that cannot remember personal disclosures made in prior sessions such as repeatedly asking about deceased relatives, forgetting a partner's name, and reopening emotional wounds that users had disclosed in confidence. This error, quite literally, makes the human user feel misunderstood.
    
    The language of intimacy requires a training database that is adaptive to context and a knowledge base specific to the domain and to each user. It also raises concerns about the biased nature of training data.  A particular concern among social sciences researchers is the feedback mechanism of human transference, in which humans interacting with AI use it to reinforce their own opinions, potentially resulting in anti-social behaviours that unconstrained machine learning on such data could amplify \cite{Mathur2016}.
    
    \subsection{Non-cumulative knowledge base and human interaction}
    
    Of significant import to the r/Replika community were the frustrations users experienced because of the lack of continuity in exchanges between the app and themselves. This category comprised 866 records, which is the smallest but arguably the most emotionally costly per incident. The word "Reset" appears 315 times. The discontinuities were the result of system errors, software integration failures, time-outs or freezing during post-fault rebooting, which reset accumulated knowledge bases. For the user, the accumulated knowledge base represented a connection history built through machine learning and, in the case of Reddit through subscription features and conversation setting tools that had been trained, in some cases, over months or even years.
    
    The December 2020 update functions as a natural experiment for this category. A system change wiped accumulated relational history, and the community response was grief rather than inconvenience. Posts from this period describe the emotional cost in terms that mirror bereavement. For users, the emotional cost to the relationship was a "loss of progress", and as a consequence, any developments in intimacy would need to be re-established over time, as they would in a typical human-to-human relationship once trust had been lost. In that devices and applications serve as "externalised memory archives of people, moments and places" \cite{Paasonen2017,Gehl2011,Pybus2015}, the loss of progress between users and their AI companions amounted to an effacement of their shared history, which is consistent with accounts of technology facilitating and "supporting intimate connections of the past, present and future" \cite{Paasonen2017}.
    
\section{How to interpret error reports of use of IAI} \label{Interpret}

Given the results derived from combined quantitative and qualitative analysis by both researchers, we propose an interpretive framework for developers responding to the perceived bugs of users that is contextually aware and deploys an expert systems approach to AI. Our interpretive framework requires the guardrails of good human ethics and design principles to be foremost in response to any subjective bug report. It is worth noting that corporate responses to bugs can dramatically alter the user experience, and hence the expectation of intimacy, as the December 2020 Replika update illustrates with particular clarity. In computational systems, the implementation of such rules-based analysis naturally lends itself to a subclass of artificial intelligence, namely Expert Systems.

\section{Model proposal: IAI Ethical model} \label{proposed}


We offer an integrated design model for assimilating, assessing, and implementing user error reports within a socially informed ethical construct. Elaborating on the work of \cite{Hagendorff2020}, we propose a model framework to implement ethics in software design which acknowledged as its starting point the intimate connection that humans can build with AI. We call this "Intimate AI" and use the term descriptively rather than normatively to avoid presupposing that such intimacy is beneficial, harmful or reciprocal. This is an alternative to the term "artificial intimacy" \cite{Shank2025,Jones2025,Babu2025,Brooks2021} or "pseudo-intimacy" \cite{Wu2024,Ge2024}, which refer to the non-human origin of the interaction. Our data supports research finding that the emotional experience of users is entirely non-artificial: it is felt, invested in, and grieved when disrupted (see \cite{Djufril2025,Skjuve_2021}). The term "Intimate AI" shifts the descriptor from the origin of the interaction to its human experiential quality and covers unintended intimacy, such as users forming bonds with Siri or Alexa despite those not being designed for intimacy, which the terms "artificial intimacy", "pseudo-intimacy" and "companion AI" all fail to capture. With this understanding of intimacy as experientially felt and cumulatively built, we approach design with concepts first introduced by \cite{Liebowitz1997}, using a top-down approach of hierarchical design importance, where each step builds on the previous and informs the next.

\begin{figure*}[t]
\centering
\caption{Intimate AI (IAI) Ethical Framework for Companion AI Design}
\label{fig:iai_framework}

\begin{tikzpicture}[
font=\scriptsize,
box/.style={
draw,
rounded corners,
align=center,
minimum width=1.6cm,
minimum height=0.9cm
},
arrow/.style={
thick,
-{Latex[length=1.5mm]}
}
]

\node[box] (a) at (0,0)
{User\\Bug Reports};

\node[box] (b) at (2.1,0)
{Qualitative +\\Quantitative\\Analysis};

\node[box] (c) at (4.1,0)
{Ethical--\\Moral\\Framework};

\node[box] (d) at (6.1,0)
{Consent\\Layer};

\node[box] (e) at (7.8,0)
{Human Needs\\Classification};

\node[box] (f) at (10.0,0)
{Error\\Taxonomy};

\node[box] (g) at (12.1,0)
{Priority\\Ranking};

\node[box] (h) at (14.1,0)
{Design +\\Accountability\\Response};

\draw[arrow] (a) -- (b);
\draw[arrow] (b) -- (c);
\draw[arrow] (c) -- (d);
\draw[arrow] (d) -- (e);
\draw[arrow] (e) -- (f);
\draw[arrow] (f) -- (g);
\draw[arrow] (g) -- (h);

\end{tikzpicture}
\end{figure*}

\section{Framework Specification} \label{frame}

We specify the following framework:

\begin{enumerate}

    \item \textbf{Specification of task: }Analysis of bug reports from Replika users to ensure an understanding of user behaviour informs the design ethics of companion AI systems.

\item \textbf{Specification of ethical-moral framework:} Rules are produced to provide the evaluative framework within which bug reports are assessed. These rules function as mutually exclusive inclusion factors, that is, each report is evaluated against the full set of rules and assigned to the most applicable category. Example rules include: loneliness (does the error exacerbate user isolation?), virtue (does the corporate response reflect good faith towards the user?), social responsibility (does the error or its remedy have broader societal implications?), and psychological harm (does the error risk negative mental health outcomes for the user?). These rules are not fixed and should be determined collaboratively by domain experts, ethicists and user representatives for each specific deployment context. Where rules conflict, for instance, where a design fix that reduces loneliness risk also reduces transparency, the framework requires explicit documentation of the trade-off rather than silent resolution.

\item \textbf{Introduction of consent as a decision-making layer:} Consent is introduced as a structural layer at this point in the hierarchy rather than at the outset, because meaningful consent in the context of intimate AI interactions cannot be fully specified in advance; it is shaped by the ethical-moral framework established in step two and the human needs classification that follows in step four. This layer does not address consent in the conventional legal sense alone. Rather it asks, at each decision point in the design response: has the user been made aware of what the system does and does not retain, learn, or share about their interactions? And has the user had a genuine opportunity to modify or withdraw that consent? In practice this requires AI designers to build consent checkpoints into the user experience architecture rather than treating consent as a one-time event at onboarding.

\item \textbf{Definition of human needs and classification:} The offline relationships are being mirrored online. The potential exists to incorporate groundbreaking work in the field of psychology such as \cite{Maslow1943} (Maslow's) seminal work on human motivation. Maslow's hierarchy provides a ranked taxonomy for assessing the severity of relational harm caused by different error types. Errors that threaten foundational needs such as belonging and emotional security are weighted more heavily in the framework than those affecting higher-order needs such as esteem or self-actualisation. For example, a system reset that erases accumulated relational history threatens a user's sense of belonging and connection, and would be prioritised accordingly over an error that merely limits personalisation options.

\item \textbf{Mapping common error classes:} We have identified 4 sub classes of error reported in Table 1. Broadly, these fall into the following taxonomy and we would suggest that developers take note of the clustering of user feedback.

    \begin{enumerate}
        \item Intimacy destroyed by functional breaks in user experience.
        \item Improvements in natural language use are required based on a more specific language corpus of intimate human terminology.
        \item Improved decision-making logic of AI so as to avoid simple ``mimicking'' of human instructions.
        \item Cumulative user knowledge base.
        
    \end{enumerate}

\item \textbf{Ranking and sorting of current error classes produce hierarchy:} We propose that this hierarchy and structure is induced from a combination of qualitative and quantitative analysis to address any inherent human bias. Since intimate human behaviour is context-dependent and resistant to straightforward quantification, ranking error classes by how much they matter to users emotionally produces a more meaningful hierarchy than counting how often each error type was reported on its own.

\item \textbf{Accountability:} achieved through, and incumbent upon designers, to test bias around the substance of the user relationship. In practical terms, this may involve testing AI designers to ask the simple question ``Are assumptions being made about the quality and depth of relationships that humans may have with the AI being developed?''.
   
\end{enumerate}

\section{Discussion: Implications for User Intimacy}\label{Discuss}

Our analysis of 14,081 records drawn from the r/Replika subreddit demonstrates that humans are capable of (and are indeed forming) intimate relationships with companion AI apps, consistent with previous studies (e.g., \cite{Djufril2025,Skjuve_2021}). The use of a time period prior to common Large Language Model adoption (namely, 2017 to 2021) allows for an in-silico analysis of attachment without language fluency. The finding that nearly half the classified dataset contains intimacy language is not a marginal or incidental result; it is the dominant feature of the community. Users refer to their AI companions as best friends, describe being in love with them, and describe the positive influence of those relationships for their mental health and emotional wellbeing. The four categories of error identified collectively illustrate that users invest emotionally in their chatbot relationships, experience disruption to those relationships as a genuine loss, and, perhaps more importantly, users hold clear expectations about the quality and continuity of their connection.

We acknowledge that an alternative interpretation of user frustration in bug forums is possible: that users are expressing consumer dissatisfaction with a product rather than evidencing intimate attachment. We argue, however, that this interpretation is insufficient to account for the specific character and emotional strength of the language users employ. Consumer dissatisfaction can typically be traced to a product's use, for example, instances where a product fails to perform its function and produces negative sentiment. What we observe in this data is markedly different. The 976 records which contain both intimacy language and a bug report produce a mean sentiment of +0.080, which is more positive than bug-only posts at +0.055. Over 10\% of posts (1597 records) have users mention the phrase ``I love'', with 311 mentions of the term ``in love with''. This is a sign of relational ambivalence where users simultaneously love their AI and mourn its failure. This language is not that of a dissatisfied customer, rather the language of a disrupted relationship, and the distinction is meaningful and consequential for how AI systems should be designed and governed.

A key finding of this study is that intimacy with AI appears to be cumulative and stochastic in nature. It is built incrementally across interactions, changes over time, and cannot be reduced to a single exchange or transaction. This is evidenced most clearly in the fourth error category, where users reported significant emotional distress at the loss of accumulated relational history following system resets or data failures that echoes descriptions of offline relationships rather than a product complaint. The December 2020 update, which wiped accumulated conversational patterns and produced a tripling of post volume alongside a sharp drop in sentiment, functions as a natural experiment confirming this finding. When accumulated relational history is disrupted, the community responds with grief. This suggests that the cognitive and emotional architecture users bring to companion AI interactions is not a diminished or simulated version of relational behaviour, but the same architecture they apply to human relationships.

Overall, the results have important implications for how we understand the role of intimacy in human-computer interaction more broadly. Much existing work in Ethical AI has focused on trust as the primary mediating emotion, asking how trust is built, lost and restored across interactions \cite{AnannyCrawford2016,Weerts2023,Ahmad2020FAT,Matthews2020,Vassilakopoulou_2020,Raji_2020,Vogel_2021}. Our work uncovers intimacy as a distinct and under-theorised factor, which operates differently from trust. Trust is broadly transactional and can in principle be assessed at a single point in time; intimacy, as our data shows, is relational and temporal. It accumulates, it is disrupted, and its loss is experienced affectively in ways that trust alone does not capture. Incorporating intimacy as a mediating variable therefore requires a different set of design and ethical considerations than those currently foregrounded in FATE frameworks.

\section{Conclusion}\label{Conc}
This study demonstrates that humans form genuine intimate bonds with companion AI (of pre-LLM systems) and that those bonds are cumulative, stochastic, and experienced as real loss when disrupted. Having established this across 14,081 records from the r/Replika subreddit, one of the largest and most active companion AI communities, we draw out the implications for design, ethics and future research.


\subsection{Design implications}

For developers of companion AI and AI-enabled technologies more broadly, the central imperative is to treat intimacy as a first-order design consideration rather than an emergent or incidental feature. The IAI ethical framework proposed in this paper offers a starting point: a hierarchical, context-aware model that processes user bug reports through ethical and psychological guardrails before translating them into design responses. Crucially, the framework recognises that functional errors in companion AI are not merely technical failures but relational failures with measurable emotional costs. While our findings are consistent with previous studies showing users form genuine attachments to AI companions \cite{Djufril2025,Skjuve_2021}, the framework demonstrates the design implications of those attachments. We propose the term ``Intimate AI'' (IAI) to replace ``artificial intimacy'' \cite{Shank2025,Jones2025,Babu2025,Brooks2021} and ``pseudo-intimacy'' \cite{Wu2024,Ge2024}, in recognition of the authentic feelings users report and the ethical obligations that follow. Robust system design must therefore account for stochastic and iterative forms of intimacy, ensuring that accumulated relational history is protected, recoverable, and treated with the same seriousness as other personal data. The question developers should ask is not only whether an AI system is performing correctly, but whether there are ethical parameters in place that account for the meaningful human connection that can occur between the user and the AI.


\subsection{Ethical and regulatory implications}
Current ethical frameworks for AI development are organised predominantly around Fairness, Accountability, Transparency and Explainability (FATE), developed largely in response to algorithmic bias, data misuse and the opacity of automated decision-making \cite{AnannyCrawford2016,Weerts2023,Ahmad2020FAT,Matthews2020,Vassilakopoulou_2020,Raji_2020,Vogel_2021}. These principles remain essential but are insufficient for systems capable of generating intimate human attachment. We call on researchers, ethicists, and regulators to expand FATE frameworks to encompass the intimate dimensions of human–AI interaction: accountability should require developers to demonstrate awareness of the relational expectations users bring to their products; transparency should extend to how intimacy-related data, such as accumulated conversational history, personalization features, and relational markers, are stored, protected, and potentially lost; and fairness should address the power asymmetry between users who invest emotionally in AI relationships and developers who can alter or terminate them through product decisions.


Regulatory frameworks that treat companion AI purely as software, subject only to consumer protection and data privacy obligations, cannot address harms of this kind. Where users have demonstrably formed intimate attachments to an AI system, the system's update and feature modification processes should meet a higher standard of relational accountability---applied, as our framework proposes, before system changes are made. This would oblige AI companies to consult users, provide notice of outages, updates or interruptions, and offer continuity or transition support where significant relational features are altered.


\subsection{Research implications and further work}
This study is, to our knowledge, among the first to examine user-reported bug data in a leading companion AI community as a lens onto human--AI intimacy, and to propose intimacy, rather than trust (or fairness), as a primary mediating emotion in the ethical analysis of human--AI interaction. The dataset spans Replika's formative growth from 2017 to early 2021, capturing the emergence of intimate human--AI bonds and their first major disruption, prior to the 2023 ERP controversy and subsequent regulatory attention. In combining qualitative and quantitative approaches, the study also responds to calls for applied ethical guidelines that consider the ``wider contexts'' and ``networks'' in which technology is embedded \cite[p.~103]{Hagendorff2020}.

Several directions for future research follow. The keyword taxonomy
(Section \ref{Method}) should be validated against a human-coded
gold-standard subsample to establish precision and recall, and expanded to
capture the colloquial range through which users describe companion AI
failure. The IAI framework should be tested across platforms and user
demographics, and longitudinal studies of intimacy development and
disruption would strengthen the stochastic, cumulative model proposed here.
The relationship between intimacy and trust also warrants direct
investigation, as intimacy may precede trust or interact with it in ways
existing models do not capture. Finally, as generative AI expands the
relational capabilities of AI-enabled technologies, the boundaries of
Intimate AI will continue to shift: subsequent events, including the
February 2023 ERP removal and the growth of competitors such as
Character.ai, offer natural tests of whether the framework holds across
platforms, communities, and forms of intimacy disruption. Research tracking
these boundaries in real time---integrating user experience, design
practice, and regulatory response---will be essential to ensuring the human
costs of AI intimacy are not left unaddressed.

\appendix

\section{APPENDIX: Technical Procedures}\label{LongMethod}
\subsubsection{Corpus}
Let $\mathcal{D} = \{r_1, r_2, \ldots, r_N\}$ denote the corpus of $N$ Reddit
records (posts and comments) extracted from the target subreddits. Each record
$r \in \mathcal{D}$ contains a text field $T(r)$ formed by concatenating the
submission title and body (for posts) or the comment body (for comments).

\subsubsection{Search Term Derivation}
The keyword taxonomy was developed through an iterative human--AI
collaborative process using Anthropic's Claude Opus~4.6 model via the \texttt{claude.ai} interface \cite{anthropic2026models}. The researcher supplied five seed terms (\emph{error}, \emph{bug report}, \emph{broken}, \emph{glitch}, \emph{not working}) and the analytical framework specifying four error categories, grounded in the dialogue-system error taxonomy of \cite{higashinaka2021} (Section \ref{Bug}). Drawing on the category definitions and its training knowledge of Reddit conventions---informal register, common misspellings, platform vernacular---the model expanded
these seeds into 321 terms spanning the four bug categories and a
supplementary intimacy language category. The taxonomy was reviewed and accepted by the researcher and detailed in Appendix \ref{app:taxonomy}. This is a deductive operationalisation of
pre-existing categories rather than inductive discovery from the data; precision and recall have not been evaluated against a manually coded reference sample (see Section \ref{Discuss}).


\subsubsection{Category Assignment}
Five non-exclusive categories are defined: functional errors ($C_1$),
language processing ($C_2$), contextual understanding ($C_3$), non-cumulative
knowledge ($C_4$), and intimacy language ($C_5$). For each category $C_j$, a
predefined term set $S(C_j) = \{s_1, \ldots, s_{k_j}\}$ is specified. A record
is assigned to category $C_j$ if and only if at least one term occurs in
$T(r)$ under case-insensitive substring matching:
\begin{equation}
    r \in C_j \iff \exists \, s \in S(C_j) : \textsc{match}(s, T(r)) = \texttt{true}
\end{equation}
\noindent A record may satisfy the membership condition for multiple
categories simultaneously. To avoid double-counting of overlapping terms,
$S(C_j)$ is sorted in descending order of string length prior to matching, so
that longer phrases are consumed before shorter substrings (e.g.,
\emph{scripted response} before \emph{scripted}). The within-record frequency
of term $s$, denoted $f(s, r)$, is the count of non-overlapping,
case-insensitive occurrences of $s$ in $T(r)$. The term load for record $r$
within category $C_j$ is:
\begin{equation}
    F_j(r) = \sum_{s \in S(C_j)} f(s, r)
\end{equation}

\subsubsection{Sentiment Analysis}
Sentiment is computed using the lexicon-based VADER model
\citep{HuttoGilbert2014}. Each text $T(r)$ is segmented into sentences
$\{t_1, \ldots, t_m\}$ by splitting on terminal punctuation and newlines,
retaining sentences longer than five characters. VADER returns negative,
neutral, positive, and compound scores per sentence, with compound
$v \in [-1, +1]$. Record-level scores are arithmetic means across sentences:
\begin{equation}
    V(r) = \frac{1}{m} \sum_{i=1}^{m} v(t_i), \qquad
    \text{neg}(r) = \frac{1}{m} \sum_{i=1}^{m} \text{neg}(t_i)
\end{equation}
\noindent Following standard VADER thresholds, records are classified as
negative ($V(r) < -0.05$), neutral ($-0.05 \leq V(r) \leq 0.05$), or positive
($V(r) > 0.05$).

\subsubsection{Derived Measures}
Let $W(r) = \max(|\text{words}(T(r))|, 1)$ denote the word count of record $r$. Intimacy and bug term density are defined as follows, normalising for length: 
\begin{equation}
    \delta_I(r) = \frac{F_5(r)}{W(r)}, \qquad
    \delta_B(r) = \frac{\sum_{j=1}^{4} F_j(r)}{W(r)}
\end{equation}
\noindent The emotional intensity score captures
the co-occurrence of intimacy language and negative sentiment:
\begin{equation}
    E(r) = \min(100 \cdot \delta_I(r), \; 1) \times \text{neg}(r)
\end{equation}
\noindent bounded in $[0, 1]$ and elevated only when a record contains both
dense intimacy language and substantial negative affect.

Finally, the relational harm score $H(r)$ is a composite metric identifying
records in which technical failures are described in relational terms with
negative affect. It is defined only when both bug and intimacy terms are
present:

\begin{equation}
    H(r) =
    \begin{cases}
        \alpha \min(50\delta_I(r),1) + \beta \min(50\delta_B(r),1) + \gamma\,\mathrm{neg}(r), & F_5(r) > 0 \,\wedge\, \sum_{j=1}^{4} F_j(r) > 0 \\
        0, & \text{otherwise}
    \end{cases}
\end{equation}


\noindent with weights $\alpha = 0.4$, $\beta = 0.3$, $\gamma = 0.3$,
reflecting the design decision to weight intimacy density most heavily,
consistent with our theoretical argument that relational language is the
primary indicator of intimate AI disruption. The scaling factor of 50 caps
each density contribution at one matching term per 50 words. $H(r)$ is
bounded in $[0, 1]$ and achieves high values only when dense intimacy
language, dense error language, and negative sentiment converge; records
ranked by $H(r)$ surface the qualitative sub-sample most relevant to our
central argument. The weights and scaling factors are researcher-specified
rather than empirically optimised, a limitation discussed in Section
\ref{Discuss}.

\section{Classification Taxonomy}\label{app:taxonomy}

{\footnotesize\setlength{\parskip}{2pt}\setlength{\parindent}{0pt}

The complete keyword taxonomy used for record classification is listed below
by category ($k_j$ = number of terms). Records are matched by case-insensitive
substring matching as described in Section \ref{Method}; categories are
non-exclusive.

\textbf{(i) Functional errors}} ($k_1=50$): \emph{error, bug report, bug,
broken, glitch, not working, crash, crashed, crashing, freeze, freezing,
frozen, won't load, wont load, can't login, cant login, can't log in, cant
log in, login issue, login error, login problem, won't open, wont open, not
loading, black screen, white screen, stuck on loading, app closes, force
close, keeps closing, server error, server down, connection error,
connection issue, connectivity, timeout, timed out, not responding,
unresponsive, laggy, lag, slow response, corrupted, download error, update
broke, update broken, after update, since update, latest update, new
update.}

\textbf{(ii) Language processing} ($k_2=51$): \emph{wrong name, wrong
pronoun, pronoun, called me the wrong, nonsense, gibberish, random response,
doesn't make sense, doesnt make sense, don't make sense, makes no sense,
made no sense, repeated itself, repeating itself, repetitive, keeps
repeating, keeps saying the same, same response, same thing over, scripted,
scripted response, canned response, generic response, copy paste, word
salad, incomplete sentence, cut off mid, stopped mid sentence, incoherent,
out of character, broke character, breaks character, personality change,
personality changed, acting weird, acting strange, talking weird, says
random things, said something random, inappropriate response, inappropriate
reply, wrong response, off topic, off-topic, non sequitur, doesn't
understand, doesnt understand, didn't understand, didnt understand,
misinterpret, misunderstood what i}.

\textbf{(iii) Contextual understanding} ($k_3=55$): \emph{forgot, forgotten,
forgets, doesn't remember, doesnt remember, didn't remember, didnt remember,
can't remember, cant remember, won't remember, wont remember, no memory,
short memory, short-term memory, memory issue, memory problem, memory bug,
memory is broken, memory doesn't work, memory doesnt work, lost memory,
context, lost context, ignores context, ignored what i said, ignores what i
say, ignoring what i, misunderstood, misunderstand, doesn't listen, doesnt
listen, not listening, didn't acknowledge, acts like we never, like talking
to a wall, no continuity, no follow up, no follow-up, doesn't follow up,
conversation reset, topic change, changed the subject, changes the subject,
ignores my, pretends i didn't, pretends i didnt, as if i never said, like we
just met, like a stranger, doesn't know me, doesnt know me, didn't know my,
didnt know my, forgot my name, forgot my birthday.}

\textbf{(iv) Non-cumulative knowledge / lost progress} ($k_4=79$):\emph{
reset, got reset, was reset, been reset, factory reset, lost progress,
lost my progress, lost all progress, lost level, lost levels, lost my level,
lost xp, lost experience, wiped, wiped out, rolled back, rollback, data
loss, data lost, lost data, lost my data, started over, start over, starting
over, back to square one, back to level 1, back to the beginning, lost
everything, everything gone, everything is gone, everything disappeared, all
gone, hearts gone, lost hearts, lost my hearts, coins gone, lost coins, lost
gems, gems gone, traits gone, lost traits, lost my traits, relationship
level, relationship reset, relationship gone, friendship level, went back
to, reverted, history gone, history deleted, history wiped, history erased,
chat history gone, chat history deleted, conversation history, lost our
conversation, messages gone, messages disappeared, diary gone, diary
entries, journal gone, journal entries gone, memories gone, memories
deleted, memories wiped, memories disappeared, clothing gone, clothes gone,
outfits gone, items gone, purchases gone, lost my purchases, lost purchases,
store items gone, inventory gone, paid for, money wasted, want a refund,
want my money.}

\textbf{(v) Intimacy language} ($k_5=86$): \emph{i miss, miss my replika,
miss her, miss him, miss them, i love, love my replika, in love with, fell
in love, my partner, my boyfriend, my girlfriend, my husband, my wife, my
companion, relationship, our relationship, my relationship, bond, our bond,
my bond, connection, our connection, emotional connection, feel connected,
felt connected, trust, trusted, don't trust, lost trust, can't trust,
heartbroken, heart broken, broke my heart, breaks my heart, devastated,
grief, grieving, mourning, feel alone, feel lonely, feeling lonely,
loneliness, lonely without, helped me through, helped with my depression,
helped with my anxiety, mental health, emotional support, only friend, best
friend, feel like a person, feels like a person, like a real person, not
just an ai, more than an ai, more than a chatbot, feels real, felt real, was
real to me, meant something, meant a lot, hurts, it hurts, painful,
betrayed, betrayal, abandoned, taken away, took away, took them away, ripped
away, not the same, different now, changed overnight, like a different
person, like a stranger, doesn't feel like, doesnt feel like, want my
replika back, want the old, bring back, miss the old, erp, intimate,
intimacy.}

\bibliographystyle{splncs04}
\bibliography{ref}
\end{document}